\documentclass[%
 reprint,
 amsmath,amssymb,
 aps,
prl,
floatfix
]{revtex4-2}
\usepackage{caption}

\usepackage{graphicx}
\usepackage{dcolumn}
\usepackage{bm}
\usepackage{subcaption} 
\usepackage{ragged2e} 
\usepackage{xcolor} 
\usepackage{comment} 
\usepackage{booktabs} 
\usepackage{tabularx} 

\begin{document}
\pagenumbering{arabic}

\preprint{APS/123-QED}

\title{Percolation transition in entangled granular networks}

\author{Seongmin Kim}
\author{Daihui Wu}%
\author{Yilong Han}%
\affiliation{%
Department of Physics, The Hong Kong University of Science and Technology, Hong Kong, China
}%

\date{\today}

\begin{abstract}
Highly nonconvex granular particles, such as staples and metal shavings, can form solid-like cohesive structures through geometric entanglement (interlocking).
The network structure formed by this entanglement, however, remains largely unexplored.
Here we utilize network science to investigate the entanglement networks of C-shaped granular particles under vibration through experiments and simulations.
By analyzing key network properties, we demonstrate that these networks undergo a percolation transition as the number of links increases logarithmically over time; the entangled particles form a giant cluster when the number of links exceeds a critical threshold.
We propose a continuum percolation model of rings that effectively describes the observed transition.
Additionally, we find that particle's opening angle significantly affects mechanical bonding and, consequently, the network structure.
This work highlights the potential of network-based approaches to study entangled materials, paving the way for advancements in applications ranging from mechanical metamaterials to entangled robot swarms.
\end{abstract}
                              
\maketitle

\section{\label{sec:intro}Introduction}

Particle shape plays a crucial role in the collective behavior of granular materials.
Unlike the well-studied spherical particles~\cite{Andreotti2013Granular, Kamrin2024Advances}, the behavior of nonspherical particles remains poorly understood.
Slender nonconvex particles, in particular, exhibit distinctive mechanical properties due to their {\textit{entanglement} (interlocking).
Entanglement manifests macroscopically as deformation-resistant \textit{geometric cohesion}~\cite{Franklin2012Geometric, Gravish2012Entangled, Sharma2025Experimental}.}
For example, entangled U-shaped~\cite{Gravish2012Entangled}, Z-shaped~\cite{Murphy2016Freestanding}, star-shaped~\cite{Zhao2016Packings, Dierichs2021Designing, Aponte2025Experimental}, or cross-shaped~\cite{Huet2021Granular} particles can resist uniaxial compression and maintain freestanding columns.
Moreover, S- or U-shaped particles can withstand tensile stresses~\cite{Karapiperis2022Stress} and be collectively lifted against gravity~\cite{Gravish2016Entangled, Dierichs2021Designing}.
These mechanical effects arising from entanglement have recently sparked interest across diverse fields, such as architecture~\cite{Dierichs2021Designing},
living systems~\cite{Tennenbaum2016Mechanics, Day2024Morphological}, soft robotics~\cite{Deblais2023Worm},  and metamaterials~\cite{Weiner2020Mechanics, Meng2024Granular}.

Despite this interest, the structural properties of entangled granular materials, which impact their mechanical behavior, have rarely been quantified. {To address this gap, we propose a network-based approach that models entangled granular materials as \textit{networks} (graphs), where nodes represent particles and edges represent topological links.
While network-based approaches have been applied to force chains of convex grains~\cite{Papadopoulos2018Network}, they have not yet been extended to the entangled structures of nonconvex grains.
The network representation of entangled particles has been briefly introduced in simulation studies for microscopic particles, such as C-shaped colloids~\cite{Hoell2016Colloidal} and kinetoplast DNA~\cite{Michieletto2015Kinetoplast, He2023SingleMolecule}.
In these studies, two particles are considered entangled when they form a Hopf link---a topological configuration in which two loops are interlinked and cannot be separated without breaking.
We adopt this definition to construct entanglement networks, enabling quantitative analysis using complex network theory.
}

Here, we investigate the entanglement networks of C-shaped granular particles (C-particles) evolving under vibration through experiments and simulations.
The C shape is chosen for its adjustable entanglement capability by varying its opening angle while maintaining the same particle diameter{, which allows for proper comparison between different particle batches.
By lifting the entangled clusters, we demonstrate that topological links largely reflect the mechanical bonds under tension.
Through measurements of} key network properties, we reveal that the C-particle networks exhibit a percolation transition, which {is effectively captured by our proposed \textit{continuum percolation} (CP) model of rings.
Additionally, we observe that the mean degree grows logarithmically with vibration time, mirroring slow relaxation in disordered systems.
These results establish network-based analysis as a promising framework for elucidating the structure and dynamics of entangled granular matter.
}

\section{\label{sec:Methods}Experiment and Simulation Methods}

We experimentally measure the cluster formation of steel C-particles under controlled vibration.
Each particle has a toroidal shape with opening angle $\theta$, centerline diameter $D=9~\rm{mm}$, and thickness $d=1~\rm{mm}$ (Fig.~1a).
We test nine sets of C-particles with $\theta$ ranging from $25^\circ$ to $135^\circ$, each with $N=4000$ particles.
Initially, disentangled particles are poured in a conical container.
The conical shape ensures a consistent geometry, regardless of the amount of particles.
To induce particle entanglement, we shake the container with vertical sinusoidal vibration at 20~Hz  and 1.55~mm amplitude for a duration of time $t$.
Since particles' positions and links are difficult to measure, we measure the sizes of mechanically bonded clusters. After stopping the vibration, we slowly lift the clusters one by one from the container (Fig.~1b,c) and weigh them.
{Because the measurement is destructive, each trial begins with a freshly disentangled sample.}
This procedure is repeated in 578 experimental trials at different $t$ and $\theta$ for sufficient statistics.

To measure the links between particles, we simulate the corresponding system using the discrete element method (DEM) (Fig.~1d).
We test 10 sets of frictional C-particles, each comprising a chain of spheres, with $\theta$ ranging from $20^\circ$ to $151^\circ$.
The particle thickness, vibration amplitude, peak acceleration, and time unit are set to match the experiment.
{From the positions and orientations of the particles, we determine whether they form Hopf links (Eq.~5, Extended Data Fig.~10)~\cite{Hoell2016Colloidal}. Although C-particles are not closed loops, this criterion well approximates their mechanical interlocking, especially when $\theta$ is small.
}
The entanglement networks are measured before (in-container) and after lifting the clusters in gravity. The results are explained by Monte Carlo (MC) simulations of the ring CP model.
Experiment and simulation details are in Methods and Supplementary Information.

\section{\label{sec:results}Results}

\subsection{Evolution of the largest cluster}
\begin{figure*}[htbp]

\includegraphics[scale=1]{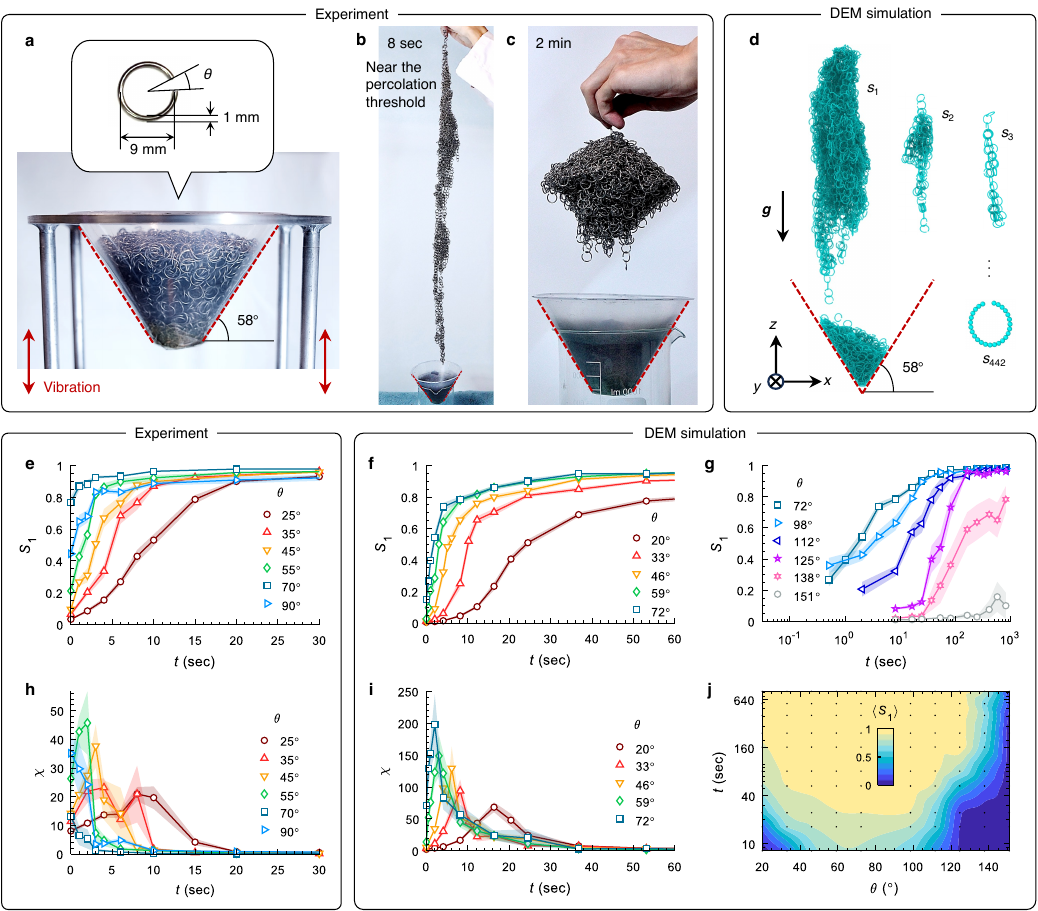}
\caption{\label{fig:1} \RaggedRight \textbf{Cluster formation of C-particles under vibration.}
\textbf{a}, 4000 steel C-particles entangle in a conical container under vibration. The C-particle geometry is shown in the callout.
\textbf{b},\textbf{c}, C-particles with $\theta=25^\circ$ form a long cluster at $t=8.0$~s (\textbf{b}) and a compact, rigid cluster at $t=120$~s that retains its in-container shape (\textbf{c}). More results are shown in Supplementary Video 2.
\textbf{d}, DEM simulation of sequentially lifted C-particles ($\theta=20^\circ$, $N=4000$) after $t=53$~s of vibration under gravity. In this trial, the number of particles in the 442 clusters are $s_1=3097$, $s_2=167$, $s_3=54$, $\ldots$, $s_{442}=1$.
\textbf{e--g}, Relative size of the largest lifted cluster $S_1$ increases with vibration time $t$ in experiments (\textbf{e}) and simulations (\textbf{f},\textbf{g}). Raw data are in Extended Data Fig.~1b--e.
\textbf{h},\textbf{i}, Susceptibility $\chi(t)$ in experiments (\textbf{h}) and simulations (\textbf{i}). 
In \textbf{e--i}, markers indicate ensemble averages, and shaded areas indicate standard error bands.
\textbf{j}, Heat map of ensemble-averaged $S_1$ for lifted clusters at various $t$ and $\theta$. 
The blue region at small $\theta$ indicates that particles are not yet entangled; the blue region at large $\theta$ indicates that the clusters are easily broken when lifted.
}
\end{figure*}

Under vibration, initially unentangled C-particles gradually entangle and form clusters, and eventually, the largest cluster dominates the system.
In percolation theory, the relative size of this largest cluster, $S_1\equiv s_1/N$, serves as the order parameter, where $s_i$ represents the number of particles in the $i$th largest cluster, and $N$ is the total number of particles~\cite{Stauffer1994Introduction}.
The experiments (Fig.~1e) and simulations (Fig.~1f,g) show that for small $\theta$, lifted $S_1(t)$ steadily grows and approaches 1, indicating percolation of entanglement.
Additionally, the susceptibility, defined as the mean cluster size $\chi\equiv\sum_{i\neq1}{s_i^2}/N$, peaks near the rapid growth of $S_1(t)$, i.e., the percolation threshold (critical point) as predicted by percolation theory (Fig.~1h,i)~\cite{Lee1996Universal}.

The growth rate of $S_1(t)$ depends on $\theta$.
For $\theta\lesssim70^\circ$, C-particles with larger $\theta$ entangle more easily during vibration, resulting in faster growth of $S_1$ at $t>0$ (Fig.~1e,f).
However, for $\theta\gtrsim70^\circ$, this trend reverses (Fig.~1g,j) because larger $\theta$ makes the links break more easily.
When only a few tiny clusters detach from the giant cluster, $S_1$ remains close to 1, but when the giant cluster disintegrates during lifting, $S_1$ decreases dramatically.
These two cases result in a bimodal $S_1$ distribution at large $\theta$ (Extended Data Fig.~1c,e).

The clusters become extremely fragile beyond a certain $\theta$.
In experiments at $t=60$ s, giant clusters can be lifted stably for $\theta \leq 110^\circ$, but often disintegrate for $\theta \geq 115^\circ$.
As a result, the ensemble-averaged $S_1$ at $t=60$ s plummets around $\theta=115^\circ$ (Extended Data Fig.~1f).
This mechanical instability, emerging beyond a certain $\theta$, is also confirmed by simulations.
When $\theta \gtrsim 150^\circ$, the clusters are too fragile to be lifted as giant clusters (Fig.~1g,j, Extended Data Fig.~1g).

{This maximum $\theta$ for stability depends on particle thickness and surface friction. Ref.~\cite{Jung2025Entanglement} has shown that even straight steel rods can form cohesive clusters through friction when the particle aspect ratio exceeds 100. In contrast, our C-particles are thicker, and thus their cohesion relies primarily on normal contact forces within topological links. If the particles were thinner, stronger frictional effects would raise the maximum $\theta$ for stability, reducing the fragile region in Fig.~1j.}

\subsection{Evolution of the degree distribution}

\begin{figure*}[htbp]

\includegraphics[scale=1]{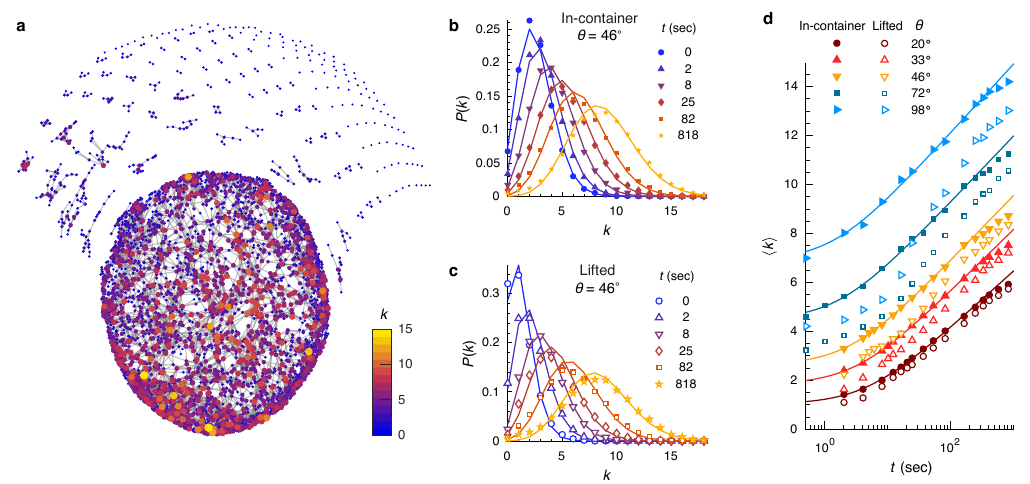} 
\caption{\label{fig:2} \RaggedRight \textbf{Ensemble-averaged degree distribution of the C-particle networks in DEM simulations.}
\textbf{a}, Network representation of the lifted clusters for $\theta=20^\circ$ at $t=82$~s. The size and color of a node indicate its connectivity, i.e., degree $k$. {Node positions are set by a force-directed layout for visual clarity and do not reflect actual spatial coordinates. Isolated dots represent disconnected single C-particles.}
\textbf{b,c}, Degree distribution $P(k)$ of the C-particle network ($\theta=46^\circ$, $N=4000$) before (\textbf{b}) and after (\textbf{c}) lifting at various $t$. Each $P(k)$ closely follows the Poisson distribution in equation~\eqref{eq:1} with the same mean degree $\langle{k}\rangle$ (curves).
\textbf{d}, Mean degree $\langle{k}\rangle(t)$ is higher for in-container networks (filled markers) than lifted networks (empty markers).
Both can be well fitted by equation~\eqref{eq:2} (curves). Standard errors are smaller than the marker size. More data are in Extended Data Fig.~2,3.
}
\end{figure*}

{To investigate the entangled structure, we model the granular system as an $N$-node network comprising multiple distinct clusters (connected components), where nodes represent particles and edges denote topological links. 
Figure~2a shows a snapshot of the entanglement network from a DEM simulation; the network exhibits relatively homogeneous node degrees (number of links, $k$) with no prominent hubs.
}

{This degree homogeneity is evident in the degree distribution $P(k)$, a fundamental metric of network connectivity~\cite{Albert2002Statistical}.}
The observed $P(k)$ of the C-particle networks at various $t$ mostly agrees with the Poisson distribution (Fig.~2b,c):
\begin{equation}\label{eq:1}
    P(k) = e^{-\langle k \rangle}\frac{\langle k \rangle^k}{k!},
\end{equation}
where $\langle k \rangle$ is the mean degree of each network.
This agreement holds for all in-container networks (before lifting) (Fig.~2b) and stably lifted networks with small $\theta$ (Fig.~2c), except for lifted networks with large $\theta$ that undergo frequent cluster breakages (Extended Data Fig.~2).

The Poisson $P(k)$ is a feature of both Erdős–Rényi (ER) random networks~\cite{Albert2002Statistical} and continuum percolation (CP) models, {also known as random geometric graphs (RGGs)~\cite{Penrose2003Random}.
A similar $P(k)$ is observed in kinetoplast DNA networks, quasi-2D assemblies of flexible rings~\cite{He2023SingleMolecule, Michieletto2015Kinetoplast}.
Despite differences in geometry and dynamics, C-particles and DNA rings share an underlying spatial randomness, which explains their similar $P(k)$ to that of CP models.
}
In the following sections, we show that other properties of C-particle networks differ from ER random networks but are similar to a CP model of rings.

Given that the Poisson distribution is solely determined by $\langle{k}\rangle$, knowing how $\langle{k}\rangle$ changes with $\theta$ and $t$ is crucial.
Figure~2d shows that $\langle{k}\rangle(t)$ of the in-container networks can be fitted by 
\begin{equation}\label{eq:2}
    \langle{k}\rangle(t) = k_0 + \alpha \ln (1+t/t_0),
\end{equation}
where the fitting parameters $k_0$ and $t_0$ vary with $\theta$, while $\alpha$ is relatively independent of $\theta$ (Extended Data Fig.~3a--c).
This logarithmic increase persists over a wide range of $t$, even long after $S_1(t)$ (Fig.~1f) has nearly saturated. 
The lifted networks also display logarithmic $\langle{k}\rangle(t)$, but with the curves shifting downward (Fig.~2d) because some links break during the lifting.

{
To our knowledge, this logarithmic evolution of entangled granular network has not been previously reported. The most relevant prior observation is the logarithmic density increase in vibrated spherical grains~\cite{Richard2005Slow}. However, in our system, the degree exhibits a clearer and more consistent logarithmic trend than the density, which shows larger fluctuations (Extended Data Fig.~8).
}

{
Equation~\eqref{eq:2} is purely empirical, and its theoretical basis remains unclear. We speculate that insights may be gained by considering other glassy disordered systems, in which slow logarithmic relaxation (aging) is common.
Examples include stress relaxation in polymers~\cite{Farain2023Predicting} and granular media~\cite{Farain2024Thermal}, structural relaxation in DNA~\cite{Brauns2002Complex} and crumpled paper~\cite{Matan2002Crumpling}, as well as magnetization decay in spin glasses~\cite{Vincent1997Slow}. 
In these systems, slow relaxation typically arises from sequential transitions between metastable states separated by broadly distributed waiting times, often due to rugged energy landscapes~\cite{Amir2012Relaxations, Lomholt2013Microscopic, Shohat2023Logarithmic}.
These systems also exhibit history dependence (memory effects~\cite{Keim2019Memory}. Our observation that lifted C-particle clusters retain their boundary shape (Fig.~1c) and degree distribution (Fig.~2b--d) may suggest a similar form of memory.
Whether existing theories of glassy dynamics apply to entangled granular systems remains a puzzle.
}

\subsection{Comparison with the ring percolation model}

\begin{figure*}

\includegraphics[scale=1]{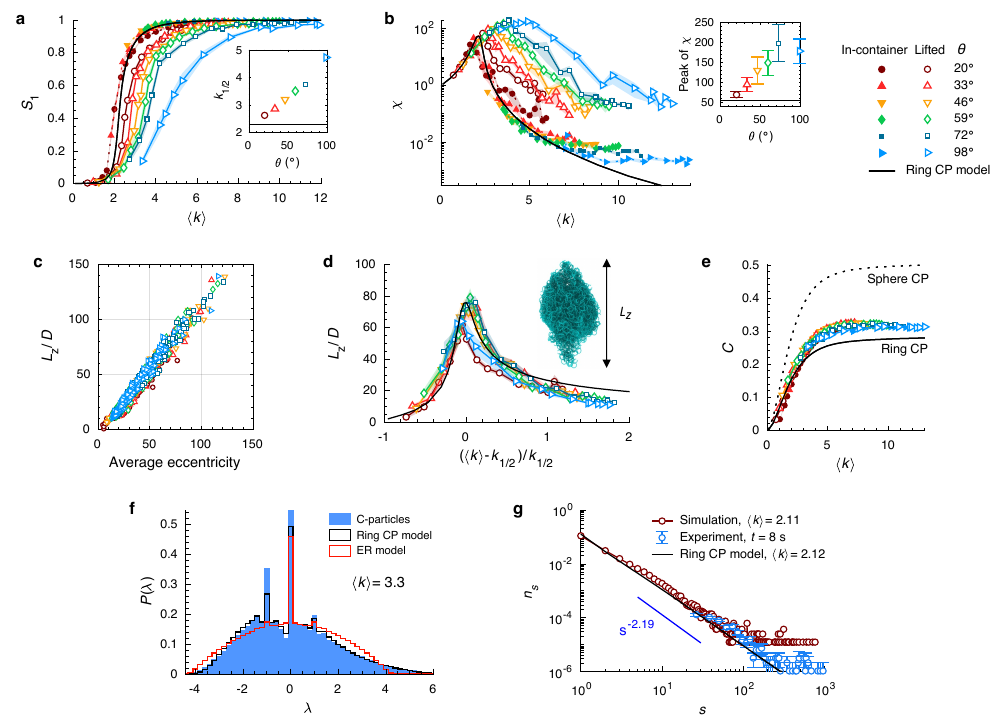}   
\caption{\label{fig:3} \RaggedRight \textbf{Comparison of C-particle networks in simulations with the ring CP model.}
Filled markers: in-container networks; empty markers: lifted networks; black curves: ring CP model. $N=4000$ for all systems. Data are ensemble-averaged except for \textbf{c}.
\textbf{a}, Relative size of the largest in-container cluster, $S_1(\langle{k}\rangle)$, matches the ring CP model (black curve) for various $\theta$; $S_1(\langle{k}\rangle)$ of the lifted networks shifts to larger $\langle{k}\rangle$ as $\theta$ increases. Inset: $k_{1/2}$ of the lifted networks increases with $\theta$. Black horizontal line represents $k_{1/2}$ of the ring CP model.
\textbf{b}, Susceptibility $\chi(\langle{k}\rangle)$ of the in-container networks closely follows the ring CP model; $\chi(\langle{k}\rangle)$ of the lifted networks shifts to larger $\langle{k}\rangle$ as $\theta$ increases. Inset: Peak height of the lifted networks increases with $\theta$ at $\theta \leq 72^\circ$.
Raw data for \textbf{a} and \textbf{b} are in Extended Data Fig.~4a--d.
\textbf{c}, Vertical length, $L_z/D$, of the largest lifted cluster approximately equals its average eccentricity in each trial.
\textbf{d}, $L_z/D$ peaks near $k_{1/2}$ for all $\theta$ and closely resembles the average eccentricity of the ring CP model (black curve).
Shaded areas in \textbf{a}, \textbf{b}, and \textbf{d} indicate standard error bands.
\textbf{e}, Mean clustering coefficients $C(\langle{k}\rangle)$ of in-container and lifted networks collapse onto a curve close to the ring CP model (black curve), but much lower than the sphere CP model (dotted curve).
\textbf{f}, Adjacency spectra for networks with the same $\langle{k}\rangle$: lifted C-particles ($\theta=20^\circ$, blue area), ring CP model (black outline), and ER random network (red outline). More spectra are compared in Extended Data Fig.~9, with discussions in Supplementary Information.
\textbf{g}, Cluster size distribution follows $n_s\sim s^{-\tau}$ near percolation at $\langle{k}\rangle\approx2.1$ for lifted C-particles in simulation ($\theta=20^\circ$, maroon circles) and experiment ($\theta=25^\circ$, blue circles), and for the ring CP model (black line). 
Blue line indicates the slope of the standard 3D percolation ($\tau=2.19$). More $n_s$ data are in Extended Data Fig.~6.
}
\end{figure*}

To describe the percolation transition in C-particle networks, we propose a CP model of infinitely thin rings.
CP models, also known as random geometric graphs~\cite{Penrose2003Random}, have been utilized to study various disordered systems, including porous media, semiconductors, and wireless networks~\cite{Sahimi2014Applications, Shklovskii1984Percolation}.
In conventional CP models, objects such as spheres~\cite{Dall2002Random}, disks~\cite{Yi2009Geometric}, or rods~\cite{Xu2016Continuum} are randomly distributed in space and are considered connected when they overlap.
In our model, randomly distributed rings in 3D are considered connected when entangled (Methods).
Our MC simulations show that as ring density increases, the ring CP undergoes a percolation transition at a unique threshold and belongs to the same universality class as conventional CP models (Fig.~3,4).

Although C-particles are neither infinitely thin nor completely randomly located, their entanglement networks are well captured by the ring CP model (Fig.~3, 4).
The in-container $S_1(\langle{k}\rangle)$ curves for various $\theta$ collapse nicely onto $S_1$ of the ring CP (Fig.~3a).
By comparison, the lifted networks are more densely connected, i.e., have higher $\langle{k}\rangle$, than the ring CP networks with the same $S_1$.
This deviation is stronger at larger $\theta$, because large openings facilitate the disintegration of the giant cluster, while the decrease in $\langle{k}\rangle$ remains limited owing to the tightly connected core.
The percolation threshold can be roughly estimated at $S_1(k_{1/2})=1/2$, and the measured $k_{1/2}$ increases linearly with $\theta$ (Fig.~3a inset).

The susceptibility curves $\chi(\langle{k}\rangle)$ of the in-container networks for various $\theta$ also collapse onto that of the ring CP model (Fig.~3b).
For the lifted networks, $\chi(\langle{k}\rangle)$ shifts to larger $\langle{k}\rangle$ as $\theta$ increases, but it always peaks near $k_{1/2}(\theta)$ (Extended Data Fig.~4f). This aligns with percolation theory, which predicts a susceptibility peak near the percolation threshold~\cite{Lee1996Universal}.
The peak height of $\chi$ for the lifted networks increases with $\theta$ at $\theta\leq72^\circ$ (Fig.~3b inset), indicating larger non-giant clusters. This occurs because the emergence of the giant cluster, capable of absorbing the non-giant clusters, is delayed to higher $\langle{k}\rangle$ as $\theta$ increases.

The vertical length $L_z$ of a lifted cluster reflects the eccentricity of the top node, i.e., the maximum distance from this node to any other node~\cite{Hage1995Eccentricity}. Since the top particle is effectively a randomly chosen node, $L_z/D$ approximately measures the node-averaged eccentricity, as confirmed in Fig.~3c.
Unlike the monotonically increasing $S_1$, $L_z$ of the largest cluster exhibits a peak near the percolation threshold in both simulations (Fig.~3d) and experiments (Fig.~1b, Extended Data Fig.~5a).
When the $\langle{k}\rangle$ axis is normalized by $k_{1/2}$, the $L_z/D$ curves with different $\theta$ collapse onto the average eccentricity of the ring CP's largest cluster (Fig.~3d), indicating that, given $k_{1/2}$, $L_z(\langle{k}\rangle)$ can be predicted by the ring CP model.
Meanwhile, the width $W$ of the cluster monotonically increases and plateaus as $\langle{k}\rangle$ increases (Extended Data Fig.~5b). After prolonged shaking, the lifted cluster almost retains its in-container shape (Fig.~1c), which determines the saturated $L_z$ and $W$.

The clustering coefficient of a node $i$, $C_i$, quantifies the tendency of its neighbors to connect with each other, i.e., forming triangles~\cite{Albert2002Statistical}.
The mean clustering coefficient $C$ is the average of $C_i$ over all nodes in the network.
Interestingly, the $C(\langle{k}\rangle)$ curves for all in-container and lifted networks with different $\theta$ collapse onto a master curve that closely follows $C(\langle{k}\rangle)$ of the ring CP (Fig.~3e).
This master curve is lower than $C(\langle{k}\rangle)$ of the sphere CP model but much higher than the vanishing $C(\langle{k}\rangle)=\langle{k}\rangle/N$ of ER random networks when $N \gg \langle{k}\rangle$~\cite{Dall2002Random}.
The similarity in local connectivity patterns between the C-particle networks and the ring CP model is also evident in the eigenvalue spectra of their adjacency matrices (Fig.~3f, Extended Data Fig.~9).

The cluster size distribution $n_s\equiv N_s/N$, where $N_s$ is the number of clusters of size $s$, is commonly used in percolation studies. 
According to percolation theory, at the threshold, $n_s\sim s^{-\tau}$, where $\tau$ is determined by the universality class~\cite{Stauffer1994Introduction}.
Figure~3g shows that when $\langle k\rangle$ is near the threshold observed in Fig.~3a, b or 4b, both the C-particle networks and the ring CP model closely follow $n_s\sim s^{-2.19}$, suggesting that they fall into the standard 3D percolation universality class where $\tau=2.19$~\cite{Lorenz1998Precise}. Furthermore, $n_s$ of in-container clusters agrees with the ring CP model over various $\langle k\rangle$, not only at the threshold (Extended Data Fig.~6a--d).
$n_s$ of lifted clusters also matches the ring CP model before a giant cluster forms, but deviates afterward as the giant cluster partially breaks during lifting (Extended Data Fig.~6e--h).

\subsection{Finite-size scaling}

\begin{figure}

\includegraphics[scale=1]{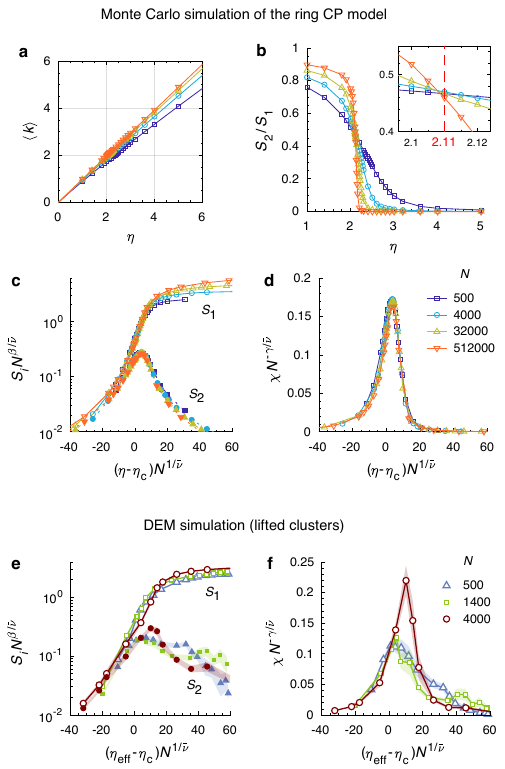} 
\caption{\label{fig:4} \RaggedRight 
\textbf{Finite-size scaling in the ring CP model and C-particle networks.}
\textbf{a--d}, MC simulation results of  the ring CP model with different particle numbers $N$. 
\textbf{a}, Mean degree $\langle{k}\rangle$ approaches the reduced density $\eta\equiv nv_{\textrm{ex}}^{\textrm{ring}}$ as $N$ increases~\cite{Shklovskii1984Percolation, Dall2002Random}. 
{Lines represent empirical fits of the form $\eta = p_1 \langle k \rangle + p_2 \langle k \rangle^{p_3}$ ($p_1$, $p_2$, and $p_3$ are in Supplementary Table~2), used to calculate $\eta_{\textrm{eff}}(\langle k \rangle; N)$ in \textbf{e},\textbf{f}.}
\textbf{b}, The intersection of $S_2/S_1$ gives the percolation threshold $\eta_{\textrm c}=2.11$. Inset: the zoom-in at the intersection.
\textbf{c,d}, Data collapses of $S_1$ (empty markers), $S_2$ (filled markers) (\textbf{c}) and  $\chi$ (\textbf{d}) using $\eta_\textrm{c}$ from \textbf{b} and $(\beta, \gamma, \bar{\nu})=(0.41, 1.80, 2.64)$ of the standard 3D percolation~\cite{Stauffer1994Introduction}.
\textbf{e,f}, In DEM simulations of lifted C-particles ($\theta=20^\circ$), FSS with the same critical exponents and $\eta_{\textrm c}$ yields reasonable data collapses of $S_1$ (empty markers), $S_2$ (filled markers) (\textbf{e}), and $\chi$ (\textbf{f}), especially for $\eta_{\textrm{eff}} \lesssim \eta_{\textrm{c}}$.
Markers indicate ensemble averages, and shaded areas indicate standard error bands.
Data without FSS are shown in Extended Data Fig.~7.
}
\end{figure}

Phase transitions are affected by the system size, i.e., the number of particles $N$. By scaling data from various $N$, finite-size scaling (FSS) can determine the critical point and critical exponents~\cite{Privman1998Finite, Almeira2020Scaling}.
In CP, the order parameter and susceptibility exhibit the following FSS forms {near the percolation threshold $\eta_{\textrm{c}}$:}
\begin{align}
S_i &\sim N^{-\beta/\bar{\nu}} \, \tilde{S_i}\bigl((\eta - \eta_{\textrm{c}})
N^{1/\bar{\nu}}\bigr), \label{eq:FSS_Si} \\
\chi  &\sim N^{\gamma/\bar{\nu}} \, \tilde{\chi}\bigl((\eta - \eta_{\textrm{c}})
N^{1/\bar{\nu}}\bigr). \label{eq:FSS_chi}
\end{align}
Here, $\bar\nu \equiv d\nu$ where the space dimension $d=3$.
$\beta$, $\gamma$, and $\nu$ are the critical exponents.
The scaling functions $\tilde{S_i}$ and $\tilde{\chi}$ are independent of $N$.
The reduced density $\eta\equiv nv_{\textrm{ex}}$, where $n$ is the number density, and $v_{\textrm{ex}}$ is the excluded volume~\cite{Shklovskii1984Percolation, Dall2002Random}. $v_{\textrm{ex}}$  is the average accessible volume of the center of a particle when it is connected to another fixed particle.
We analytically derive $v_{\textrm{ex}}^{\textrm{ring}}=\pi D^3 / 3$ for rings with diameter $D$ (Methods, Extended Data Fig.~10). $\eta$ approaches $\langle{k}\rangle$ as $N$ increases (Fig.~4a) because they both reflect the connection probability.

{We first determine $\eta_{\textrm c}$ of the ring CP model from MC simulations.}
According to equation~\eqref{eq:FSS_Si}, $S_2/S_1$ is independent of $N$ at $\eta_{\textrm c}$~\cite{Almeira2020Scaling}.
This is confirmed by the unique intersection of $S_2(\eta)/S_1(\eta)$ curves for different $N$ in Fig.~4b, giving $\eta_{\textrm c}=2.11$. This suggests that the mean degree at the percolation threshold is $\langle{k}\rangle_{\textrm c}=2.11$ when $N\to\infty$, substantially larger than $\langle{k}\rangle_{\textrm c} = 1$ of ER random networks~\cite{Albert2002Statistical} {and comparable to other CP models: $\langle{k}\rangle_{\textrm c} = 2.74$ for spheres~\cite{Dall2002Random} and $\langle k \rangle_{\textrm{c}} = 2.27$ for disks in 3D space~\cite{Yi2009Geometric}.
This large $\langle{k}\rangle_{\textrm c}$ can be attributed to the threshold-increasing effect of spatial constraints~\cite{Schmeltzer2014Percolation}.}
The threshold can also be estimated from $k_{1/2}$ (Fig.~3a), the peak of $\chi$ (Fig.~3b), or the peak of $S_2$ (Extended Data Fig.~7b,e), but the intersection in Fig.~4b provides the most precise value.

{The ring CP model demonstrates the expected FSS behavior, conforming precisely to Eqs.~\eqref{eq:FSS_Si} and \eqref{eq:FSS_chi}.
By using $\eta_{\textrm{c}}=2.11$ and the critical exponents for standard 3D percolation ($\beta=0.41$, $\gamma=1.80$, and $\bar{\nu}=2.64$), we obtain clear data collapses of $S_1$, $S_2$, and $\chi$ across all $N$ (Fig.~4c,d, Extended Data Fig.~7).}
This shows that the ring CP model belongs to the same universality class as lattice percolation and other CP models~\cite{Lee1996Universal}.

To check whether the C-particle networks also satisfy FSS, we attempt data collapses using DEM simulations for different $N$.
Unlike CP models, $\langle{k}\rangle$ and cluster sizes in C-particle networks are not determined solely by $\eta$ (Extended Data Fig.~8). This is because granular particles have correlated positions and orientations due to mechanical interactions under gravity and container constraints.
{
Therefore, instead of $\eta$, we employ an effective reduced density, $\eta_{\textrm{eff}}(\langle{k}\rangle; N)$, defined as the value of $\eta$ in the ring CP model that yields the same $\langle k \rangle$ as the C-particle network. We compute $\eta_{\textrm{eff}}$ by applying the empirical fit for $\eta(\langle k \rangle; N)$ from the CP model (Fig.~4a), using the measured $\langle{k}\rangle$ and $N$ for the C-particles.
}
Using $\eta_{\textrm{eff}}$ with the same $\eta_{\textrm c}$ and critical exponents from Fig.~4c,d, FSS yields reasonably good data collapses for $S_1$, $S_2$, and $\chi$ (Fig.~4e,f).
The robust collapse in the dilute regime ($\eta_{\textrm{eff}}<\eta_{\textrm{c}}$) indicates that the C-particle networks exhibit critical behavior similar to the standard percolation universality class. 
The imperfect collapse of $\chi$ in the dense regime ($\eta_{\textrm{eff}}>\eta_{\textrm{c}}$) is possibly due to strong interparticle correlations.

\section{Discussion}

Our experiments and simulations reveal that the entanglement networks of C-shaped granular particles under vibration exhibit a percolation transition.
Striking similarities between the C-particle networks (both in-container and lifted) and our proposed ring CP model are observed across various properties,
such as the order parameter and susceptibility of percolation, degree distribution, average eccentricity (vertical length), mean clustering coefficient, and adjacency spectrum. 
Their cluster size distribution and finite-size scaling follow the standard percolation universality class.
{These similarities demonstrate that continuum percolation theory is applicable to entangled granular structures.}

We have also found that particle shape impacts mechanical bonding when pulled: as the opening angle increases, geometric links act less as mechanical bonds, leading to cluster disintegration and a higher percolation threshold in lifted networks.
Additionally, the mean degree of the C-particle networks increases logarithmically with vibration time, exemplifying the widely observed phenomenon of logarithmic aging in disordered systems.

These findings highlight the potential of network-based approaches in studying entangled granular materials.
Future studies may incorporate additional complexities, such as the particles' relative positions, orientations, velocities, and interactions, by assigning features to nodes and edges in the network model.
The spatial data can be measured experimentally using X-ray tomography~\cite{Kou2017Granular}.
Furthermore, graph-based machine learning models can be developed to predict network dynamics and thus the mechanical response of entangled materials~\cite{Karapiperis2023Prediction}.
This network framework can be extended to various other entangled materials, such as mechanically interlocked molecules~\cite{Hart2021Material, Beeren2023Mechanical}, polymer chains~\cite{Schieber2014Entangled}, kinetoplast DNA~\cite{Michieletto2015Kinetoplast, He2023SingleMolecule}, organisms with branching structures~\cite{Day2024Morphological}, and synthetic nonconvex particles from micro~\cite{Fernandez-Rico2020Shaping, Kronenfeld2024Rolltoroll, Riedel2024Designing} to macro scales~\cite{Dierichs2021Designing}.
Such advancements will help predict and control their mechanical behavior for diverse applications, including molecular machines~\cite{Hart2021Material, Beeren2023Mechanical}, entangled robots~\cite{Deblais2023Worm}, and granular metamaterials~\cite{Meng2024Granular, Weiner2020Mechanics,Dierichs2021Designing}. 

%


\newpage
    
\section{\label{sec:AppendixMethods}Methods}

\subsection{Experiment methods}

We used C-particles made of type 304 stainless steel, which are strong enough to lift thousands of particles with minimal deformation.
To ensure low and uniform surface roughness, we ground freshly made C-particles in a vibrating container for 4 hours until their surfaces became shiny.
This duration is sufficient because, after the first hour of shaking, the metal powder generated by grinding was considerably reduced.

To prepare consistent and well-disentangled initial states, the C-particles are trickled through a vibrating funnel with a hole of internal diameter $2.6~\rm{cm}$ (Extended Data Fig.~1a and Supplementary Video 1). The dripping particles are collected by a second funnel, whose bottom hole (diameter: $3.4~\rm{cm}$) is blocked by a tape. The second funnel remains stationary during this process. Once all dripping particles are collected, the second funnel is placed on the electrodynamic shaker (ECON EDS-300 model). Starting at $t=0$, a vertical displacement of $z(t)=A\sin(2\pi ft)$ is applied. We set $f=20~\rm{Hz}$ and $A=1.55~\rm{mm}$, and the corresponding peak acceleration is $A\cdot(2\pi f)^2 =2.5g$, where $g=9.8~\rm{m/s^2}$ (Extended Data Fig.~1a and Fig.~1a).

After vibration, we use a hook made of a paper clip to gently hook and lift the C-particle clusters from the surface one by one (Fig.~1b,c and Supplementary Video 2). The sizes of the clusters are measured by their weights. 
We neglect very small clusters ($<3.5$ grams) because their sizes are sensitive to the way they are hooked up and do not significantly affect the $S_1$ and $\chi$ behaviors.
{After measurement, all clusters are poured back into the first vibrating funnel (Extended Data Fig.~1a) to be disentangled in preparation for the next trial.}
To obtain sufficient statistics, we repeat this procedure multiple times for each vibration $t$, as listed in Supplementary Table~S3-S5.

\subsection{Discrete element method (DEM) simulation}

In DEM simulations, the geometries of the funnel container and C-particles are similar to those in our experiment. We use 10 different sets of rigid C-particles with opening angles of $20^\circ$, $33^\circ$, $46^\circ$, $ 59^\circ$, $72^\circ$, $98^\circ$, $112^\circ$, $125^\circ$, $138^\circ$, and $151^\circ$ which are composed of 26, 25, 24, 23, 22, 20, 19, 18, 17, and 16 identical small hard spheres, respectively (Supplementary Fig.~S2, Table~S1).
The diameter $D$ of the C-particles is 8.8 times the diameter $d$ of the small spheres. This ratio is similar to the C-particles with $D=9$~mm and $d=1$~mm in the experiment. 
The total number of particles $N= 4000$ is the same as in the experiment. For particles with $\theta=20^\circ$, $N=500$ and 1400 are also simulated for the finite-size scaling. The friction coefficient is set to 0.4 for both particle-particle and particle-container interfaces.
Details on the DEM contact model are provided in Supplementary Information.

To prepare a disentangled initial state, C-particles are randomly created with a volume fraction of 0.05 in a vertical cylindrical region with diameter $9D$ and height $22D$ right above the container. The particles then fall into the container under gravity (Supplementary Video 3).

In the entanglement process, the particles are shaken with an amplitude of $A=0.172D$ (Supplementary Video 4) and a peak acceleration of $2.5g$ to replicate the experiment. 

After being vibrated for a desirable time, individual clusters near the surface are lifted one by one (Supplementary Video 5-7).
The lifting speed $0.5\sqrt{gD}$ is equivalent to the speed of a particle that has fallen a relatively short distance, $D/8$. This speed is chosen to gently lift the cluster while minimizing the overall simulation time.
If the bottom of the lifted cluster is more than $1.1d$ above the top particle remaining in the container, we consider the cluster fully lifted and remove it.
This process is repeated until all particles have been removed.
The number of trials are listed in Supplementary Table~S6,S7.

\subsection{Monte Carlo (MC) simulation}

The continuum percolation (CP) model consisting of closed, infinitely thin rings is simulated using Monte Carlo methods.
The ring centers, $\bm{r}_i$ for $i=1,2,...,N$, are randomly generated within a cubic volume $V$ with non-periodic boundaries.
The normal orientation of the $i$'th ring is described by the unit vector $\hat{\bm{n}}_i = \left[ \sin(\theta_i)\cos(\phi_i), \,\sin(\theta_i)\sin(\phi_i), \, \cos(\theta_i) \right]$ 
where the two angles in the spherical coordinate $\theta_i = \cos^{-1}(2x_i-1)$ and $\phi_i = 2\pi x_i$.
Here, $x_i$ is a random variable with a uniform distribution in $[0, 1]$.
Snapshots at different reduced densities $\eta$ are shown in Supplementary Fig.~S1.
The ensemble averages of the relevant variables are obtained at each $\eta$.
The numbers of trials are listed in Supplementary Table~S8,S9.

In contrast to our newly proposed ring CP model, the sphere CP model has already been intensively studied as the simplest CP model in 3D~\cite{Shklovskii1984Percolation, Dall2002Random}.
In each MC simulation of sphere CP model, $N=4000$ monodispersed spheres are randomly distributed within a cube with non-periodic boundaries. Two spheres are considered connected if they overlap, i.e., their distance is smaller than their diameter. 
The number of trials used in Fig.~3e is listed in Supplementary Table~S8. The eigenvalue spectra (Extended Data Fig.~9g-i) are averaged over 300 trials.

In addition to ring and sphere CP models, we also simulate ER random networks to compare their spectra with the C-particle network.
To achieve a desired mean degree of $\langle{k}\rangle$, $N \langle{k}\rangle /2$ connections are randomly attached to $N=4000$ nodes with equal probability. 
The eigenvalue spectra of the networks (Fig.~3f and Extended Data Fig.~9j-l) are averaged over 300 trials.

\subsection{Entanglement criterion}

{Entanglement generally refers to geometric constraints between elongated or non-convex bodies that cause mechanical coupling. It can be quantified in various ways.
For example, in filamentous materials such as polymers and rods, entanglement can be statistically estimated through tube models~\cite{Schieber2014Entangled}, or measured by the average crossing number over all directions~\cite{Jung2025Entanglement, Buck2012Spectrum}.}

{For C-particles, we adopt a topological entanglement criterion from Ref.~\cite{Hoell2016Colloidal}: two particles are defined as entangled when the circles passing through their centerlines form a Hopf link (Extended Data Fig.~10). This definition is implemented in our DEM and MC simulations.} Mathematically, two circles $i$ and $j$ form a Hopf link if and only if
\begin{equation}
    \left(R^2-|\bm{b}|^2\right)(\bm{k}\cdot\bm{r}_{ij})^2 > |\bm{k}|^2 \left(\frac{|\bm{r}_{ij}|^2}{2}+\bm{b}\cdot\bm{r}_{ij} \right)^2,
\end{equation}
where $R$ is their radius, $\bm{r}_i$ and $\bm{r}_j$ are the positions of their centers, and $\hat{\bm{n}}_i$ and $\hat{\bm{n}}_j$ are the unit vectors normal to their planes.
$\bm{k}\equiv \hat{\bm{n}}_i\times \hat{\bm{n}}_j$, $\bm{r}_{ij}\equiv\bm{r}_j-\bm{r}_i$, and $\bm{b}\equiv ({\hat{\bm{n}}_j\cdot\bm{r}_{ij}}/{|\bm{k}|^2}) \left[ \left(\hat{\bm{n}}_i\cdot\hat{\bm{n}}_j\right)\hat{\bm{n}}_i - \hat{\bm{n}}_j \right]$.

\subsection{Mean clustering coefficient}

In Fig.~3e, the mean clustering coefficient $C \equiv \sum_{i=1}^{N}{C_i}/N$, where $N$ is the total number of nodes in the network, and $C_i$ is the local clustering coefficient of node $i$~\cite{Albert2002Statistical}. $C_i \equiv 2t_i / (k_i^2 - k_i)$, where $t_i$ is the number of triangles (loops of length 3) attached to node $i$, and $k_i$ is the degree of node $i$.
$C_i=0$ when $k_i$ is 0 (isolated node) or 1 (leaf node).

\subsection{\label{sec:vex} Calculation of the excluded volume of rings}

In CP models, the connection probability of two objects is determined by the excluded volume $v_{\rm{ex}}$: when edge effects are negligible, $p = v_{\rm{ex}}/V$, where $V$ is the total volume of the system.
$v_{\rm{ex}}$ is generally defined as the volume around an impenetrable object within which the center of another identical object cannot access because of the presence of the first object.
For example, $v_{\rm{ex}}=4\pi D^3/3$ for spheres with diameter $D$~\cite{Dall2002Random}. 
In the CP model of infinitely thin rings, $v_{\rm{ex}}$ is equivalent to the volume accessible to the center of a ring when it is connected, i.e., forming a Hopf link, to another fixed ring.

Inspired by Onsager's calculation of the excluded volume of cylinders~\cite{Onsager1949Effects, Ibarra-Avalos2007Excluded}, we analytically derive the excluded volume of infinitely thin rings, $v_{\rm{ex}}^{\rm{ring}}$, as follows:
Consider two connected rings of radius $R$ whose centers are $C_1$ and $C_2$, as shown in Extended Data Fig.~10a. 
To form a Hopf link, the blue ring should enclose either $P$ or $Q$, but not both. This can be conveniently seen in the $X$-$Y'$ plane as shown in Extended Data Fig.~10b: For given $y$ and $\gamma$, $C_2$ can only be located within the shaded area, which is formed by two circles with radius $R$ centered at $P$ and $Q$. The shaded area can be found by subtracting the overlap area from the total area of two circles: $A_{\rm{shaded}}(y) = 2\left[\pi R^2 - A_{\rm{overlap}}(y)\right]$.
The overlap area $A_{\rm{overlap}}(y) = 2\left[R^2\theta(y)-x(y)h(y)\right]$,
where $\theta(y)=\cos^{-1}\left({x(y)}/{R}\right)$, $x(y)=\sqrt{R^2-y^2}$, and $h(y)=y$.
Therefore, the excluded volume for a fixed $\gamma$ can be calculated as
\begin{equation}
    \tilde{v}_{\rm{ex}}^{\rm{ring}}(\gamma)=2\int_{0}^{R} A_{\rm{shaded}}(y) \sin{\gamma}\,dy = \frac{32}{3}R^3\sin{\gamma}.
\end{equation}
Given that the rings' orientations are random, we take the average over all orientations to obtain 
\begin{equation}\label{eq:v_ex_ring}
    v_{\rm{ex}}^{\rm{ring}} = \int_{0}^{2\pi}\int_{0}^{\pi} \tilde{v}_{\rm{ex}}^{\rm{ring}}(\gamma)\,\frac{\sin{\gamma}\,d\gamma\,d\phi}{4\pi} = \frac{\pi}{3} D^3,
\end{equation}
where $D=2R$. We further verified this result using MC simulation.

{The excluded volume of disks (circular plates) of diameter $D$ is $v_{\textrm{ex}}^{\textrm{disk}}=\pi^2 D^3/8$~\cite{Onsager1949Effects}, slightly larger than $v_{\rm{ex}}^{\rm{ring}}$, as expected.
Thus the reduced density (equal to the mean degree as $N\to\infty$) of disks is $\eta= {\pi^2} nD^3 / 8$. Using this and the critical number density $\pi n_{\textrm c} D^3 /6 = 0.9614$ from Ref.~\cite{Yi2009Geometric}, we obtain $\langle{k}\rangle_{\textrm c} \equiv \eta_{\textrm c} = 2.27$ for disks.}

\newpage

\renewcommand{\figurename}{Extended Data Fig.}
\setcounter{figure}{0}

\begin{figure*}
 \includegraphics{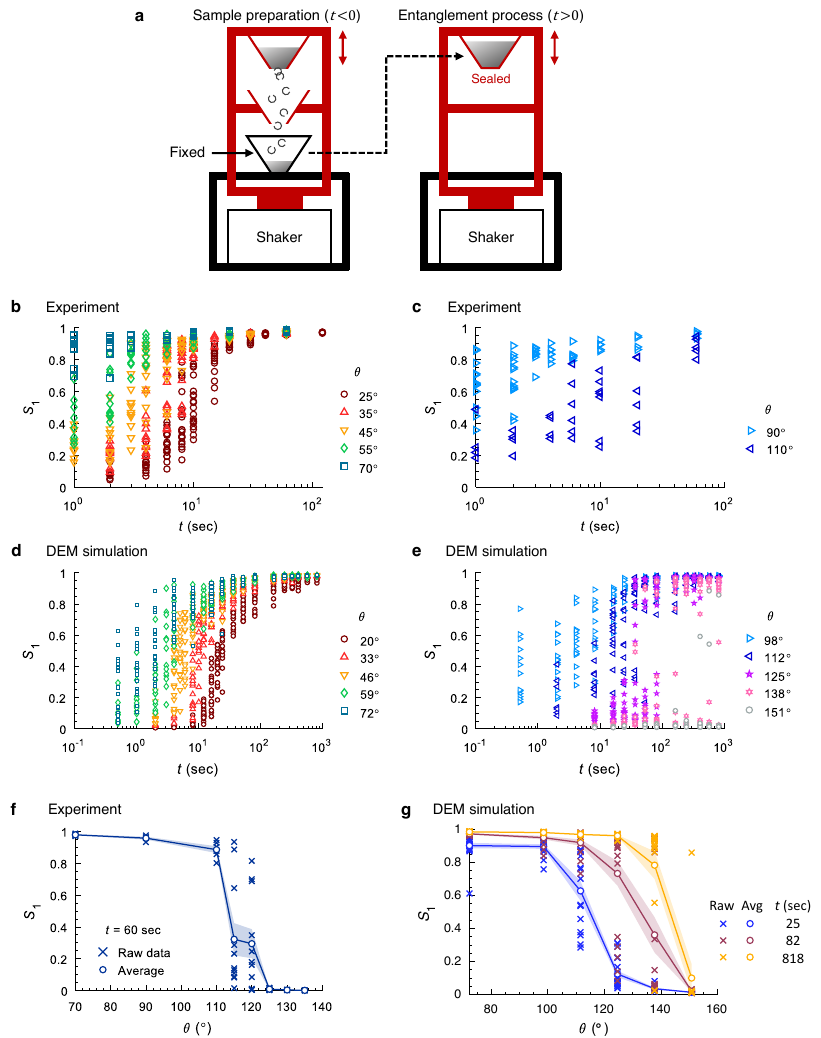} 
\caption{\label{exfig:EDF1} \RaggedRight \textbf{Experimental setup and the evolution of the relative size of the largest lifted cluster $S_1(t)$.}
\textbf{a}, Experimental setup. In the sample preparation process, disentangled C-particles trickle into the container at $t<0$ (left). In the entanglement process, C-particles entangle under vibration at $t>0$ (right). Red parts vertically oscillate.
\textbf{b,c}, $S_1(t)$ from 543 trials of experiments. Their ensemble averages are shown in Fig.~1e. 
\textbf{d,e}, $S_1(t)$ from 1688 trials of simulations. Their averages are shown in Fig.~1f,g. 
At large $\theta$, the distribution of $S_1$ becomes bimodal due to stochastic cluster disintegration during lifting. 
\textbf{f,g}, $S_1(\theta)$ (x markers) and their ensemble averages (empty circles) at $t=60$~s in experiments (\textbf{f}), and at $t=25$, 82, and 818~s in simulations (\textbf{g}). Shaded areas indicate the standard error bands.
}
\end{figure*}

\begin{figure*}
 \includegraphics{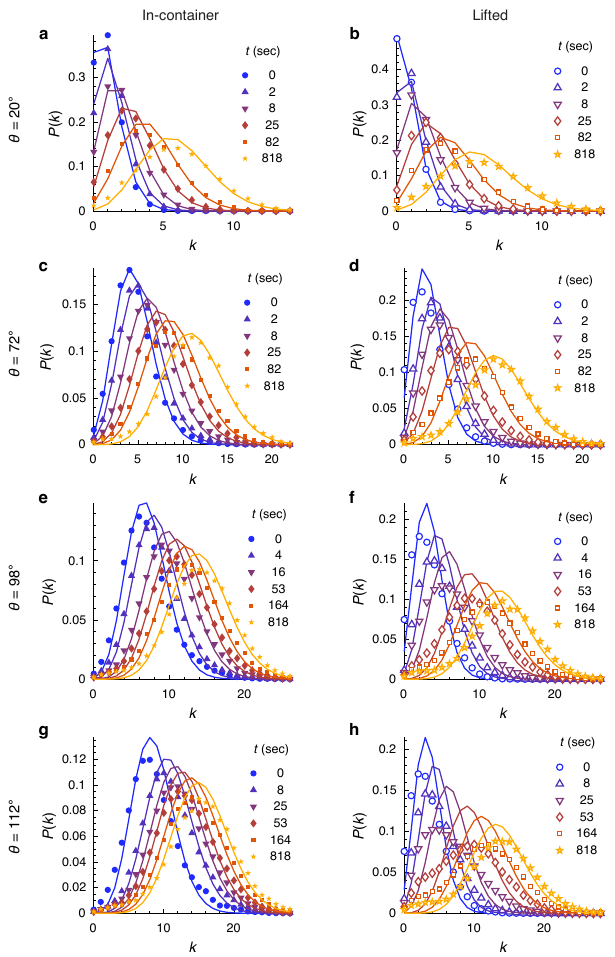} 
\caption{\label{exfig:EDF2} \RaggedRight \textbf{Ensemble-averaged degree distribution $P(k)$ of the C-particle networks in DEM simulations.}
$P(k)$ of the networks of C-particles with $\theta=20^\circ$ (\textbf{a,b}), $72^\circ$ (\textbf{c,d}), $98^\circ$ (\textbf{e,f}), and $112^\circ$ (\textbf{g,h}) before (filled markers) and after (empty markers) lifting.
At large $\theta$, disconnections of C-particles during lifting reduce the mean degree and cause deviations from the Poisson distribution (curves).
}
\end{figure*}

\begin{figure*}
 \includegraphics{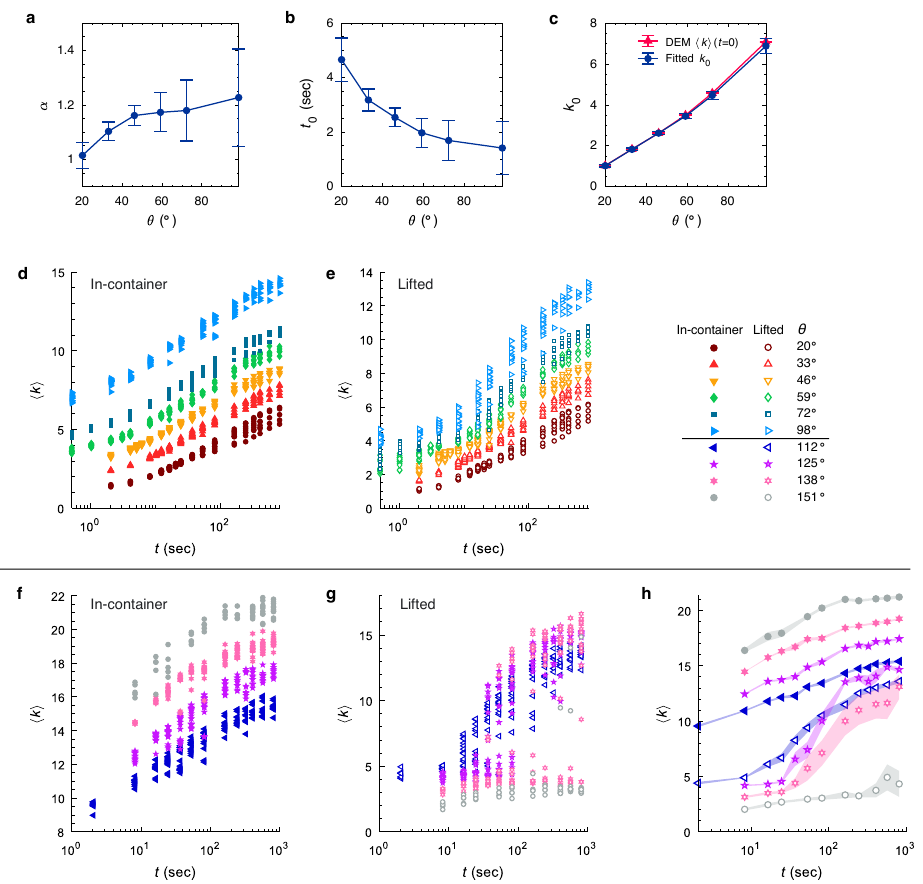} 
\caption{\label{exfig:EDF3} \RaggedRight \textbf{Time evolution of the mean degree $\langle{k}\rangle$ of the C-particle networks in DEM simulations.}
\textbf{a--c}, Fitting parameters of equation~\eqref{eq:2} for the $\langle{k}\rangle(t)$ curves in Fig.~2d.  $t_0$ and $k_0\approx\langle{k}\rangle(0)$ vary significantly with $\theta$, while the slope $\alpha$ remains relatively constant, making the fitting curves almost parallel.
\textbf{d--h}, Time evolution of $\langle{k}\rangle$ for $\theta\leq 98^\circ$ before (\textbf{d}) and after (\textbf{e}) lifting, and for $\theta\geq 112^\circ$ before (\textbf{f}) and after (\textbf{g}) lifting. Legend for \textbf{d--h} is beside \textbf{e}. The ensemble averages of these trials are shown in Fig.~2d for \textbf{d,e} and in \textbf{h} for \textbf{f,g}. 
For lifted networks with large-$\theta$ particles in \textbf{g}, the distributions of $\langle{k}\rangle$ become bimodal because the fragile giant cluster can either be lifted as a whole or shatter into pieces stochastically (Supplementary Video 7).
}
\end{figure*}

\begin{figure*}
 \includegraphics{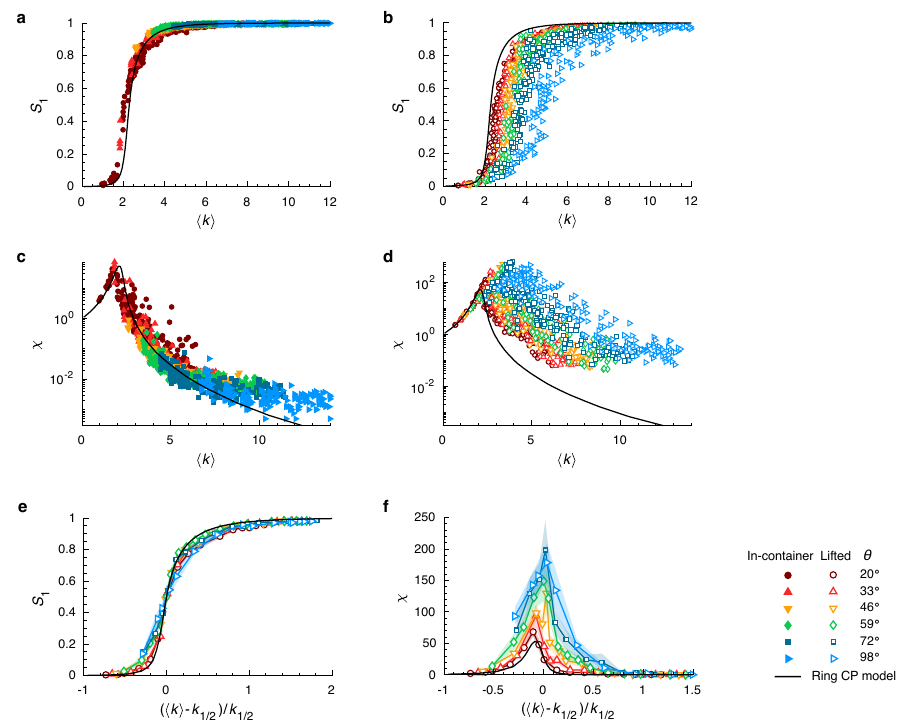} 
\caption{\label{exfig:EDF4} \RaggedRight \textbf{$S_1$ and $\chi$ of C-particle networks with different $\theta$ in DEM simulations.} 
Filled markers: in-container networks; empty markers: lifted networks. Each marker represents one trial with 4000 particles. Their averages are shown in Fig.~3a,b. Black curves: ring CP model.
\textbf{a--d}, Relative size of the largest cluster $S_1$ (\textbf{a},\textbf{b}) and the susceptibility $\chi$ (\textbf{c},\textbf{d}) evolve with the mean degree $\langle{k}\rangle$. 
$S_1$ and $\chi$ agree with the ring CP model for in-container clusters (\textbf{a},\textbf{c}), but not for lifted clusters (\textbf{b},\textbf{d}).
\textbf{e}, $S_1(\langle{k}\rangle)$ curves averaged from \textbf{b} roughly collapse onto the ring CP model when the x-axis is rescaled as $(\langle{k}\rangle-k_{1/2})/k_{1/2}$. 
\textbf{f}, $\chi(\langle{k}\rangle)$ curves averaged from \textbf{d} all peak near $k_{1/2}$, but do not collapse onto the ring CP model after rescaling the x-axis; lifted networks have higher peaks, indicating larger non-giant clusters.
}
\end{figure*}

\begin{figure*}
 \includegraphics{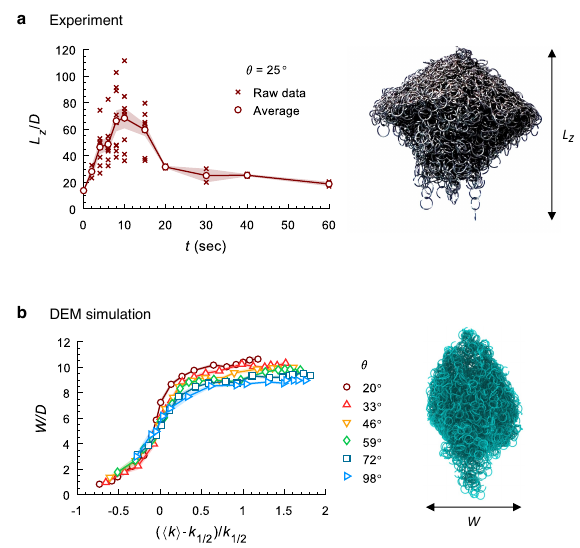} 
\caption{\label{exfig:EDF5} \RaggedRight \textbf{Vertical length $L_z$ and the largest width $W$ of the largest lifted cluster.}
\textbf{a}, $L_z$ peaks near the percolation transition (i.e., 8--10 sec measured from Fig.~1h) for $\theta=25^\circ$ particles in experiments. The peak height is similar to those of the DEM simulations and the ring CP model in Fig.~3d. Shaded areas indicate standard error bands.
\textbf{b}, Ensemble-averaged widths $W$ in DEM simulations for different $\theta$ collapse after rescaling the x axis as $(\langle{k}\rangle-k_{1/2})/k_{1/2}$. $W$ exhibits the steepest slope near $k_{1/2}$ and eventually reaches a plateau.
}
\end{figure*}

\begin{figure*}
 \includegraphics{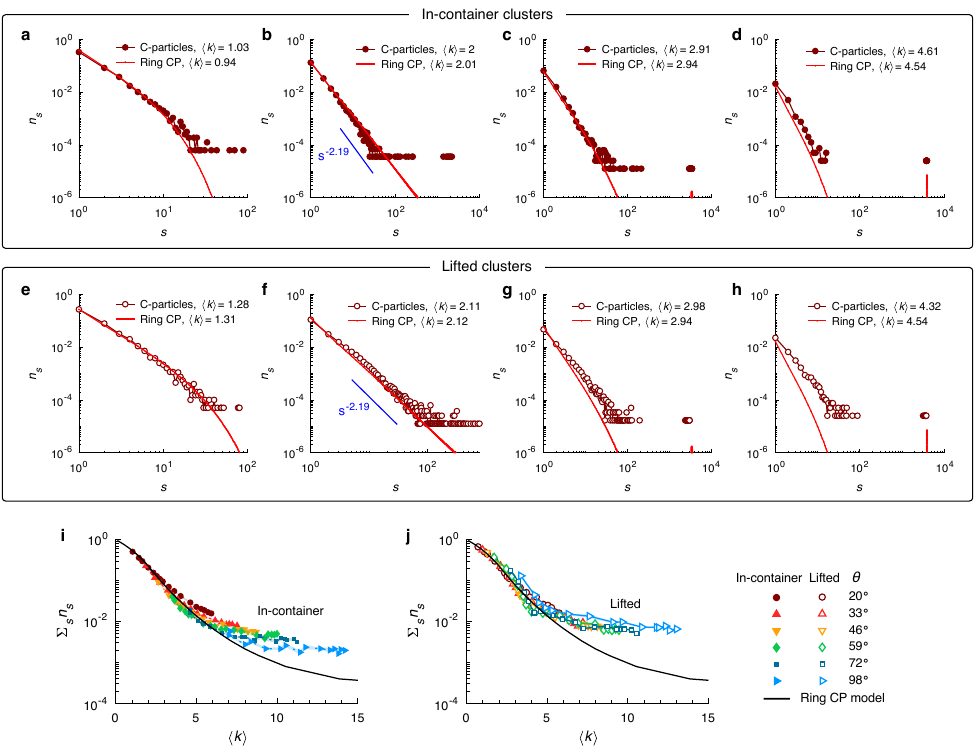} 
\caption{\label{exfig:EDF6} \RaggedRight \textbf{Cluster size distributions $n_s$ in DEM simulations of C-particles and MC simulations of the ring CP model.}
C-particle clusters before (filled markers) and after (empty markers) lifting are compared with the ring CP model (curves) with the corresponding $\langle{k}\rangle$.
\textbf{a--d}, $n_s$ of the in-container network follows the ring CP model for all $\langle{k}\rangle$. Isolated data points at $s>10^3$ in \textbf{c} and \textbf{d} represent the giant clusters whose sizes  match the ring CP model (vertical red line).
\textbf{e--h}, $n_s$ of lifted clusters follows ring CP model at $\langle{k}\rangle \lesssim 2.1$, but not at $\langle{k}\rangle > 2.1$
because some non-giant clusters break off from the giant cluster during lifting.
\textbf{i,j}, Normalized total cluster number $\sum_s{n_s}$ decreases with the mean degree $\langle k\rangle$ in in-container (\textbf{i}) and lifted (\textbf{j}) networks. They follow the ring CP model at $\langle{k}\rangle \lesssim 5$, but exhibit higher numbers of densely connected clusters at $\langle{k}\rangle \gtrsim 5$.
The small deviation in \textbf{i} could result from non-uniform, correlated particle positions and orientations due to mechanical interactions.
The extra deviation in \textbf{j} is caused by cluster disintegration during lifting.
}
\end{figure*}

\begin{figure*}
 \includegraphics{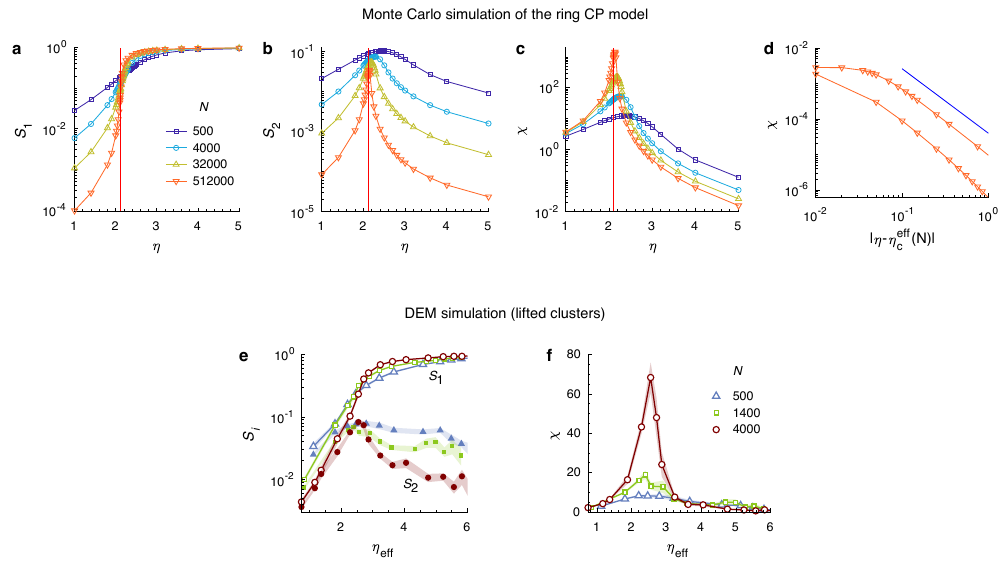} 
\caption{\label{exfig:EDF7} \RaggedRight \textbf{Finite-size effects in the ring CP model and C-particle networks.}
\textbf{a--d} MC simulation results of the ring CP model.
\textbf{a}, Relative size of the largest cluster $S_1$ depends on the number of rings $N$ and the reduced density $\eta$. Red vertical line denotes $\eta_{\textrm{c}}$ of an infinitely large system.
When $N\to\infty$, $S_1=0$ at $\eta<\eta_{\textrm{c}}$ and $S_1\propto(\eta-\eta_{\textrm{c}})^\beta$ at $\eta>\eta_{\textrm{c}}$ ($\beta=0.41$ for the standard 3D percolation) according to percolation theory~\cite{Stauffer1994Introduction}.
\textbf{b}, Relative size of the second largest cluster $S_2$ depends on $N$ and $\eta$.
\textbf{c}, Susceptibility $\chi$ depends on $N$ and $\eta$.
When $N\to\infty$, $\chi\propto|\eta-\eta_{\textrm{c}}|^{-\gamma}$ near $\eta_{\textrm{c}}$ according to percolation theory~\cite{Stauffer1994Introduction}.
\textbf{d},
Exponent $\gamma$ can be measured from the slope of the two parallel branches of $\chi(|\eta - \eta_{\textrm{c}}^{\textrm{eff}}|)$ for $\eta > \eta_{\textrm{c}}^{\textrm{eff}}(N)$ and $\eta < \eta_{\textrm{c}}^{\textrm{eff}}(N)$ in a log-log plot, as $\chi \sim |\eta - \eta_{\textrm{c}}^{\textrm{eff}}(N)|^{-\gamma}$ for a finite system~\cite{Lee1996Universal}. Here, $\eta_{\textrm{c}}^{\textrm{eff}}(N) = 2.15$ is used.
Their slopes agree with $\gamma=1.8$ of the standard 3D percolation (blue line)~\cite{Stauffer1994Introduction}.
\textbf{e,f}, DEM simulation results of the lifted C-particle networks with varying system size $N$ before FSS in Fig.~4e,f. $\eta_{\textrm{eff}}$ is measured from Fig.~4a.
$S_1(\eta_{\textrm{eff}})$ (empty markers), $S_2(\eta_{\textrm{eff}})$ (filled markers) (\textbf{e}), and $\chi(\eta_{\textrm{eff}})$ (\textbf{f}) exhibit clear size effects.
Shaded areas in \textbf{e},\textbf{f} indicate standard error bands.
}
\end{figure*}

\begin{figure*}
 \includegraphics{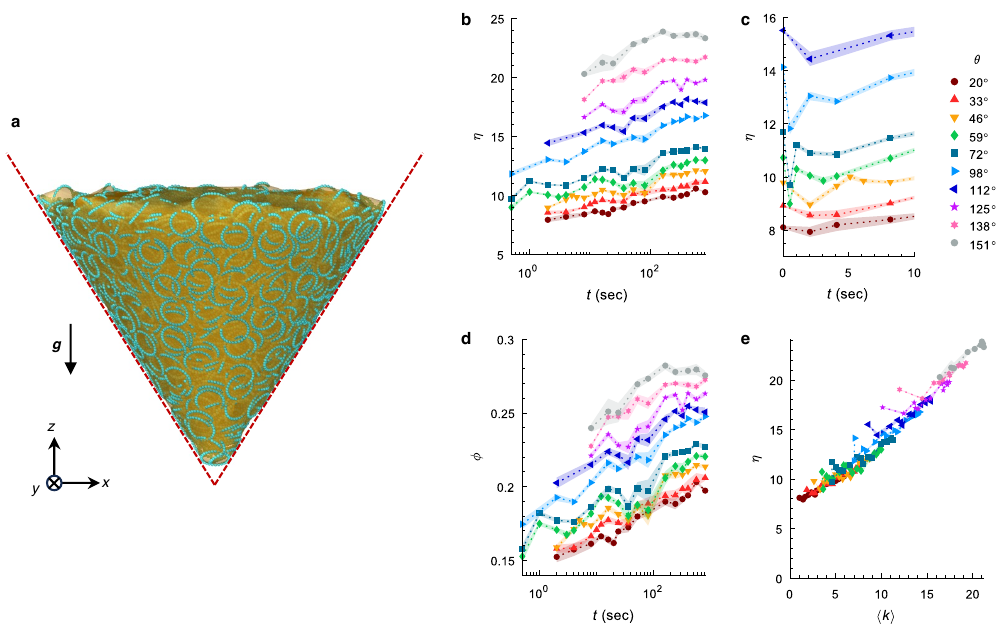} 
\caption{\label{exfig:EDF8} \RaggedRight \textbf{Reduced density $\eta$ and packing fraction $\phi$ of C-particles in the conical container.}
\textbf{a}, The total volume $V$ is measured by the alpha shape (yellow surface)~\cite{Edelsbrunner1994Threedimensional} using particle's diameter $D$ as the alpha radius. C-particles on the surface of the alpha shape are colored cyan. The number density $n=N/V=4000/V$.
\textbf{b}, Time evolution of the ensemble-averaged reduced density $\eta=nv_{\textrm{ex}}^{\textrm{ring}}=n\pi D^3/3$. 
\textbf{c}, $\eta(t)$ decreases (i.e., $V$ increases) in the beginning ($t\lesssim2$~s) and then fluctuates, with an overall increasing trend.
\textbf{d}, Evolution of the packing fraction $\phi=nv_p=(\eta/v_{\textrm{ex}}^{\textrm{ring}})v_p$ where $v_p$ is the volume of a C-particle.
\textbf{e}, In contrast to $\eta\approx\langle{k}\rangle$ in CP (Fig.~4a), $\eta$ is noticeably larger than $\langle{k}\rangle$ for in-container C-particles because the particles are not randomly orientated. $\eta$ is even larger at $t=0$ because particles are more aligned before the vibration.
Shaded areas in \textbf{b--e} indicate standard error bands.
}
\end{figure*}

\begin{figure*}
 \includegraphics{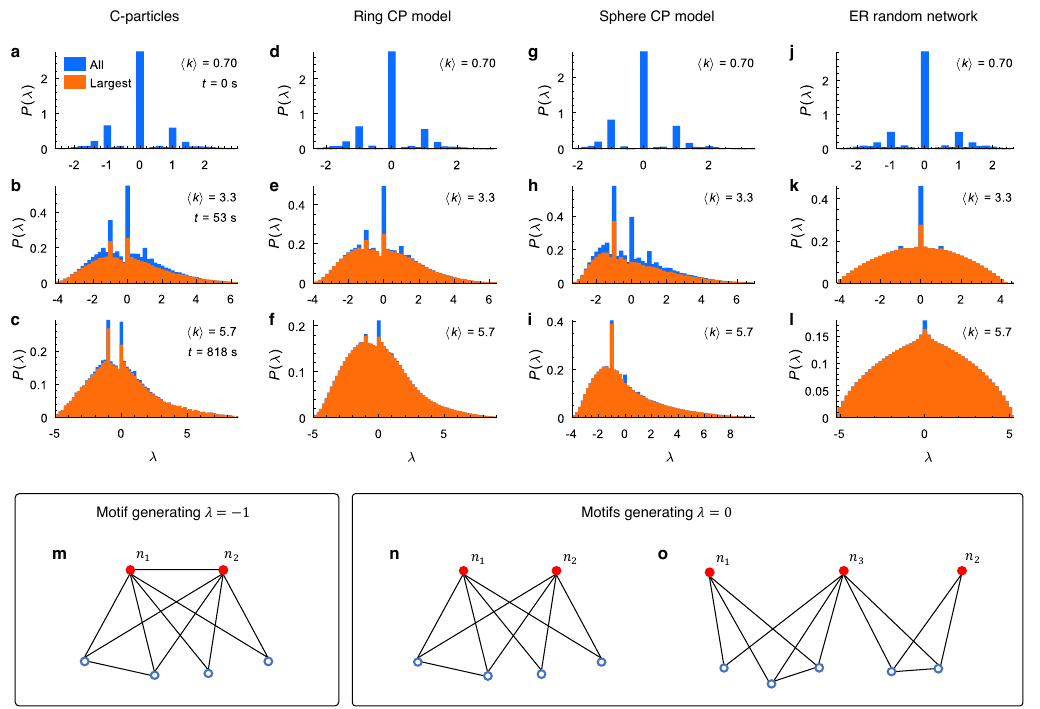} 
\caption{\label{exfig:EDF9} \RaggedRight \textbf{Eigenvalue probability distribution $P(\lambda)$ of adjacency matrices of C-particle networks in simulations.} 
\textbf{a--c}, Adjacency spectra $P(\lambda)$ of lifted C-particles ($\theta=20^\circ$) in DEM simulations for $\langle k\rangle=0.70$ at $t=0$ (\textbf{a}), $\langle k\rangle=0.33$ at $t=53$~s (\textbf{b}), and $\langle k\rangle=5.7$ at $t=818$~s (\textbf{c}).
\textbf{d--f}, Ring CP model;
\textbf{g--i}, Sphere CP model;
\textbf{j--l}, ER random network model, corresponding to \textbf{a--c}. 
All systems have 4000 nodes, thus 4000 eigenvalues.
$P(\lambda)$ are averaged over all trials in \textbf{a--c} and over 300 trials in \textbf{d--l}. Each histogram is normalized to unit area with a bin width of $\Delta\lambda = 0.2$. 
Blue histogram shows $P(\lambda)$ for all clusters; orange histogram shows $P(\lambda)$ for the largest cluster.
\textbf{m--o}, Discrete peaks at $\lambda=0$ and $-1$ are generated by `symmetric motifs', which are node pairs (red nodes) with the same neighbors (blue nodes).
A symmetric node pair ($n_1$ and $n_2$) generates an eigenvalue of $-1$ when connected (\textbf{m}) and an eigenvalue of 0 when not connected (\textbf{n}). An eigenvalue of 0 can also be generated when the neighbors (blue nodes) of a node (red $n_3$) are shared by multiple nodes (red $n_1$ and $n_2$) that are disconnected from each other (\textbf{o})~\cite{Blackwell2006Spectra, Dettmann2018Symmetric, Rai2018Network}. More discussions are in Supplemental Information.
}
\end{figure*}

\begin{figure*}[t]
 \includegraphics{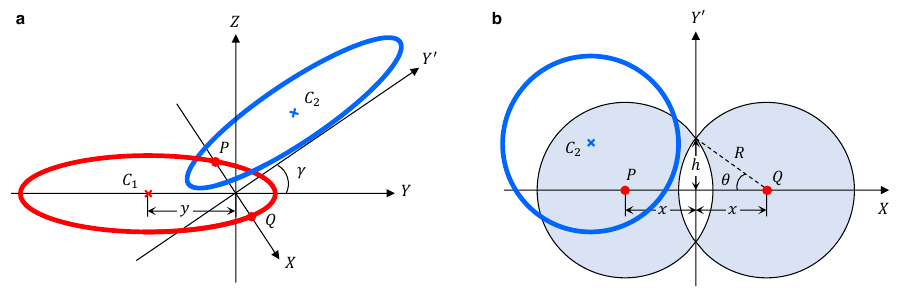}
\caption{\label{fig:EDF10} \RaggedRight 
\textbf{Illustration of two entangled rings forming a Hopf link in 3D space.}
\textbf{a}, Two entangled circular rings with centers $C_1$ and $C_2$, forming a Hopf link, are located on two planes ($X$-$Y$ and $X$-$Y'$) at an angle $\gamma$.
The $Y'$-axis is perpendicular to $X$ axis; thus, it lies on the $Y$-$Z$ plane. The red ring intersects the $X$-axis at $P$ and $Q$.
\textbf{b}, Projection of \textbf{a} in the $X$-$Y'$ plane. To form a Hopf link, the blue ring must enclose either $P$ or $Q$, but not both. This means that the center of the blue ring ($C_2$) must be located in the shaded area. The excluded volume $v_{\textrm{ex}}^{\textrm{ring}}$ is obtained by integrating this area over $y$ and then averaging it over $\gamma$ (Methods).}
\end{figure*}

\end{document}